# An exact power series expansion formula for the Källén-Sabry vacuum polarization potential

**Antonio R Martines**
*Liton Microtecnologia e Microfabricação LTDA., São Paulo, SP, Brazil*

**Abstract**

In this work it is shown that the formula for the Källén-Sabry vacuum polarization potential [1, 2] can be rewritten as a sum of integrals similar to the integral formulas for the Bickley-Naylor functions [8, 9] and for the modified second-type Bessel functions [13]. These integrals can be solved for their respective power series expansions resulting in an exact power series expansion for the Källén-Sabry vacuum polarization potential.

In the Appendix A it is also presented an explicit power series expansion for the Uehling vacuum polarization potential [1, 3, 6, 7].

## 1. Introduction

The Källén-Sabry vacuum polarization potential corresponds to the $\alpha^2 Z\alpha$ order correction for the potential generated by a static point charge Z ($e = 1$), calculated using the Källén-Sabry [1, 2] expression for the complete vacuum polarization function $\Pi(p^2)$ to second order in $\alpha$.

The Källén-Sabry potential as well as the Uehling potential [1, 3] and the Wichmann-Kroll potential [1, 4] corresponds to the lowest order quantum electrodynamics nonlinear corrections for the Coulomb potential generated by a static point charge. These potentials are needed for several applications, including the calculation of exotic and electronic energy levels [1, 15-20], and the calculation of some charged particles scattering cross sections [14]. Due to the importance of these potentials they deserve some interest in order to obtain their analytic power series formulas and this is the subject of this work, more specifically for the Källén-Sabry potential.

In a recent work, Pain [8] pointed out that, in 1972, Klarsfeld [6] suceeded to find an analytic expression for the Uehling potential in terms of the spherical Bessel functions providing a generalized expression for the Pauli and Rose formula [7]. Such expression, written in terms of Bickley-Naylor functions [8, 9], was rediscovered in 1997 by Balantekin et al. [14] and also in 2011 by Frolov and Wardlaw [10, 11], who followed a different procedure.

In this work we follow a similar procedure to that used by Klarsfeld [6] in order to convert the Källen-Sabry potential formula (2.1)-(2.2) into a more mathematically suitable expression, the result is then solved for its several component integrals, thus providing an exact power series expansion formula.

For instance, in the Appendix A it is presented the explicit Pauli and Rose [6, 7, 10, 14] power series expansion formula for the Uehling vacuum polarization potential written in terms of the modified second-type Bessel functions $K_0(x)$, $K_1(x)$ and the integral of $K_0(x)$ [13].

## 2. Calculation

The formula for the Källén-Sabry vacuum polarization potential is given by the following expression [1] :

$$V_{KS}(r) = \frac{\alpha^2 Z\alpha}{\pi^2 r} I_{KS}(2\,r) \tag{2.1}$$

where :

$$\begin{aligned}
I_{KS}(x) = \int_1^\infty e^{-x\cdot t} \Bigg\{ &\left( \frac{13}{54} t^{-2} + \frac{7}{108} t^{-4} + \frac{2}{9} t^{-6} \right)(t^2 - 1)^{1/2} + \left( -\frac{44}{9} t^{-1} + \frac{2}{3} t^{-3} + \frac{5}{4} t^{-5} + \frac{2}{9} t^{-7} \right) \ln\left(t + (t^2-1)^{1/2}\right) \\
&+ \left( \frac{4}{3} t^{-2} + \frac{2}{3} t^{-4} \right)(t^2-1)^{1/2} \ln\left(8\,t\,(t^2-1)\right) \\
&+ \left( -\frac{8}{3} t^{-1} + \frac{2}{3} t^{-5} \right)\left( \int_t^\infty \frac{(3y^2 - 1)}{y(y^2-1)} \ln\left(y + (y^2-1)^{1/2}\right) - \frac{1}{(y^2-1)^{1/2}} \ln\left(8\,y\,(y^2-1)\right) dy \right) \Bigg\} dt
\end{aligned} \tag{2.2}$$

In eq. (2.1), the value for the electron charge is set $e = 1$ and the value for the distance radius "$r$" from the point charge is given in units of reduced electron Compton wavelenght :  $\lambdabar_e = $ 386.159 fm

The above formula for the Källén-Sabry integral, $I_{KS}(x)$, can be rewritten by means of a suitable change of variables ( $t \to cosh(w)$ ) which results a sum of integrals with similar expressions for the integral formulas for the modified second-type Bessel functions [13] and to the Bickley-Naylor functions [8, 9]. It can also be shown that all the integral formulas in the resulting expression for $I_{ks}(x)$ obeys to recursion relations similar to the Sonin-Nielsen [12] recursion relations for the Bessel functions, which allows to reduce the expression for $I_{KS}(x)$ to its shortest form. Each of these recursion relations can be related to corresponding differential equations which, by their turn, allows the calculation of power series expansions for the integrals in eq. (2.2).

The resulting expressiom for $I_{KS}(x)$ has the form: $I_{ks}(x) = \sum p_j(x).F_j(x)$ , where $p_j(x)$ are polynomials and $f_j(x)$ are the functions listed in the following section. This generic expression for $I_{KS}(x)$ has the same form of the Pauli and Rose result for the integral in the formula of the Uehling potential, a sum of polynomials multiplied by special functions.

## 3. Results

The resulting power series expression for $I_{KS}(x)$ is :

$$
\begin{aligned}
I_{ks}(x) = {} & x\left(-\frac{91}{135} - \frac{x^2}{36} - \frac{4}{3}(\gamma + \ln(x/2))\right) f01(x) \\
& + \left(\frac{47}{27} + \frac{47 x^2}{6480} + \frac{x^4}{1620} - \frac{1}{12}(24 + x^2)(\gamma + \ln(x/2))\right) f02(x) \\
& + x\left(-\frac{65}{648} + \frac{419 x^2}{12960} - \frac{x^4}{3240} + \frac{1}{24}(36 + x^2)(\gamma + \ln(x/2))\right) f03(x) \\
& + \left(-\frac{41}{9} + \frac{2 x^2}{3} - \frac{5 x^4}{864} + \frac{x^6}{3240} + \frac{1}{24}(96 - x^4)(\gamma + \ln(x/2))\right) f04(x) \\
& + \frac{1}{18}(24 + x^2) f05(x) - \frac{1}{36} x(36 + x^2) f06(x) + \frac{8}{9} x\, f07(x) - \frac{1}{18}(6 + x^2) f08(x) \\
& + \frac{1}{18} x(2 + x^2) f09(x) - \frac{4}{3} x\, f10(x) + \frac{1}{36}(-96 + x^4) f11(x) - \frac{1}{18}(-96 + x^4) f12(x) \\
& - \frac{5}{72}(-96 + x^4) f13(x) + \frac{1}{144} \pi^2 (-96 + x^4) f14(x) \\
& - \frac{1}{144} \pi^2 (-6 + 2x - x^2 + x^3) f15(x)
\end{aligned}
$$
(3.0)

where :

$$
f01(x) = \int_0^\infty e^{-x.cosh(w)} \frac{1}{cosh(w)} dw = \int_x^\infty K_0(y) dy =
$$

$$
= \frac{\pi}{2} + (\gamma + \ln(x/2)) \sum_{j=0}^\infty \frac{1}{(2j+1) 2^{2j} (j!)^2} x^{2j+1} - \sum_{j=0}^\infty \frac{1}{(2j+1)^2 2^{2j} (j!)^2} x^{2j+1} - \sum_{j=1}^\infty \frac{1}{(2j+1) 2^{2j} (j!)^2} (\psi_{(j+1)} - \psi_1) x^{2j+1}
$$
(3.1)

$$
f02(x) = \int_0^\infty e^{-x.cosh(w)} dw = K_0(x) = -(\gamma + \ln(x/2)) \sum_{j=0}^\infty \frac{1}{2^{2j} (j!)^2} x^{2j} + \sum_{j=1}^\infty \frac{1}{2^{2j} (j!)^2} (\psi_{(j+1)} - \psi_1) x^{2j}
$$
(3.2)

$$f_{03}(x) = \int_0^\infty e^{-x \cdot \cosh(w)} \cosh(w)\, dw = K_1(x) =$$

$$= \frac{1}{x} + (\gamma + \ln(x/2)) \sum_{j=0}^\infty \frac{1}{2^{2j+1}\, j!\,(j+1)!} x^{2j+1} - \sum_{j=1}^\infty \frac{1}{2^{2j+2}\, j!\,(j+1)!} \left(\psi_{(j+1)} - \psi_1\right) x^{2j+1} - \sum_{j=0}^\infty \frac{1}{2^{2j+2}\, j!\,(j+1)!} \left(\psi_{(j+2)} - \psi_1\right) x^{2j+1}$$

(3.3)

$$f_{04}(x) = \int_0^\infty e^{-x \cdot \cosh(w)}\, w\, \frac{\sinh(w)}{\cosh(w)}\, dw =$$

$$= \frac{\pi^2}{24} + \frac{1}{2} (\gamma + \ln(x/2))^2 + (\gamma + \ln(x/2)) \sum_{j=1}^\infty \frac{1}{2^{2j+1}\, j\, (j!)^2} x^{2j} - \sum_{j=1}^\infty \frac{1}{2^{2j+2}\, j^2\, (j!)^2} x^{2j} - \sum_{j=1}^\infty \frac{1}{2^{2j+1}\, j\, (j!)^2} \left(\psi_{(j+1)} - \psi_1\right) x^{2j}$$

(3.4)

$$f_{05}(x) = \int_0^\infty e^{-x \cdot \cosh(w)} \ln\left(\frac{\cosh(w)}{\sinh(w)}\right) dw =$$

$$= \frac{\pi^2}{8} \sum_{j=0}^\infty \frac{1}{2^{2j}\,(j!)^2} x^{2j} - \frac{\pi}{2} \sum_{j=0}^\infty \frac{1}{((2j+1)!!)^2} x^{2j+1} + (\gamma + \ln(x/2)) \sum_{j=0}^\infty c_{1,j}\, x^{2j+2} + \sum_{j=0}^\infty c_{2,j}\, x^{2j+2}$$

$$c_{1,j} = -\frac{1}{4} \sum_{m=0}^j \frac{(2j - 4m + 1)}{(2m+1)(j-m+1)\, 2^{2m}\, (m!)^2\, ((2j - 2m + 1)!!)^2}$$

$$c_{2,j} = \frac{1}{2} \sum_{m=0}^j \frac{1}{(2m+1)\, 2^{2m}\, (m!)^2\, ((2j-2m+1)!!)^2} \left( \frac{1}{(2m+1)} + \frac{(2j-4m+1)}{2(j-m+1)} \left(\psi_{(m+1)} - \psi_1\right) \right)$$

(3.5)

$$f_{06}(x) = \int_0^\infty e^{-x \cdot \cosh(w)} \cosh(w) \ln\left(\frac{\cosh(w)}{\sinh(w)}\right) dw =$$

$$= -\frac{\pi^2}{8} \sum_{j=0}^\infty \frac{1}{2^{2j+1}\, j!\,(j+1)!} x^{2j+1} + \frac{\pi}{2} \sum_{j=0}^\infty \frac{1}{(2j+1)!!\,(2j-1)!!} x^{2j}$$

$$- (\gamma + \ln(x/2)) \sum_{j=0}^\infty (2j+2)\, c_{1,j}\, x^{2j+1} - \sum_{j=0}^\infty \left(c_{1,j} + (2j+2)\, c_{2,j}\right) x^{2j+1}$$

$$c_{1,j} = -\frac{1}{4} \sum_{m=0}^j \frac{(2j - 4m + 1)}{(2m+1)(j-m+1)\, 2^{2m}\, (m!)^2\, ((2j - 2m + 1)!!)^2}$$

$$c_{2,j} = \frac{1}{2} \sum_{m=0}^j \frac{1}{(2m+1)\, 2^{2m}\, (m!)^2\, ((2j-2m+1)!!)^2} \left( \frac{1}{(2m+1)} + \frac{(2j-4m+1)}{2(j-m+1)} \left(\psi_{(m+1)} - \psi_1\right) \right)$$

(3.6)

$$f_{07}(x) = \int_0^\infty e^{-x \cosh(w)} \frac{1}{\cosh(w)} \ln\left(\frac{\cosh(w)}{\sinh(w)}\right) dw =$$

$$= \frac{\pi}{2} \ln(2) - \frac{\pi^2}{8} \sum_{j=0}^\infty \frac{1}{(2j+1) 2^{2j} (j!)^2} x^{2j+1} + \frac{\pi}{2} \sum_{j=0}^\infty \frac{1}{(2j+2)((2j+1)!!)^2} x^{2j+2}$$

$$-(\gamma + \ln(x/2)) \sum_{j=0}^\infty \frac{1}{(2j+3)} c_{1,j} x^{2j+3} + \sum_{j=0}^\infty \left(\frac{1}{(2j+3)^2} c_{1,j} - \frac{1}{(2j+3)} c_{2,j}\right) x^{2j+3}$$

$$c_{1,j} = -\frac{1}{4} \sum_{m=0}^j \frac{(2j-4m+1)}{(2m+1)(j-m+1) 2^{2m} (m!)^2 ((2j-2m+1)!!)^2}$$

$$c_{2,j} = \frac{1}{2} \sum_{m=0}^j \frac{1}{(2m+1) 2^{2m} (m!)^2 ((2j-2m+1)!!)^2} \left(\frac{1}{(2m+1)} + \frac{(2j-4m+1)}{2(j-m+1)} (\psi_{(m+1)} - \psi_1)\right)$$

(3.7)

$$f_{08}(x) = \int_0^\infty e^{-x \cosh(w)} \cosh(w) \frac{w}{\sinh(w)} dw =$$

$$= \frac{1}{2} (\gamma + \ln(x/2))^2 \sum_{j=0}^\infty \frac{1}{(2j)!} x^{2j} + (\gamma + \ln(x/2)) \sum_{j=0}^\infty \left(\frac{1}{(2j+1)!} + (2j+1) r_{1,j}\right) x^{2j}$$

$$+ \frac{\pi^2}{6} \sum_{j=0}^\infty \frac{1}{(2j)!} x^{2j} - \frac{\pi^2}{4} \sum_{j=0}^\infty \frac{1}{(2j+1)!} x^{(2j+1)} + \sum_{j=0}^\infty (r_{1,j} - (2j+1) r_{2,j}) x^{2j}$$

$$r_{1,j} = \sum_{k=1}^{2j+1} \sum_{m=0}^k \frac{(-1)^m}{k} \frac{1}{2^{(k-m)} m! (2j-k+1)!} \left(\left(\frac{k-m}{2}\right)!\right)^{-2} \frac{(1+(-1)^{(k-m)})}{2}$$

$$r_{2,j} = \sum_{k=1}^{2j+1} \sum_{m=0}^k \frac{(-1)^m}{k} \frac{1}{2^{(k-m)} m! (2j-k+1)!} \left(\left(\frac{k-m}{2}\right)!\right)^{-2} \frac{(1+(-1)^{(k-m)})}{2} \left(\frac{1}{k} + \left(\psi_{\left(\left(\frac{k-m}{2}\right)+1\right)} - \psi_1\right)\right)$$

(3.8)

$$f_{09}(x) = \int_0^\infty e^{-x \cosh(w)} \frac{w}{\sinh(w)} dw =$$

$$= -\frac{1}{2} (\gamma + \ln(x/2))^2 \sum_{j=0}^\infty \frac{1}{(2j+1)!} x^{2j+1} - (\gamma + \ln(x/2)) \sum_{j=0}^\infty r_{1,j} x^{2j+1} + \sum_{j=0}^\infty r_{2,j} x^{2j+1}$$

$$- \frac{\pi^2}{6} \sum_{j=0}^\infty \frac{1}{(2j+1)!} x^{2j+1} + \frac{\pi^2}{4} \sum_{j=0}^\infty \frac{1}{(2j)!} x^{2j}$$

$$r_{1,j} = \sum_{k=1}^{2j+1} \sum_{m=0}^k \frac{(-1)^m}{k} \frac{1}{2^{(k-m)} m! (2j-k+1)!} \left(\left(\frac{k-m}{2}\right)!\right)^{-2} \frac{(1+(-1)^{(k-m)})}{2}$$

$$r_{2,j} = \sum_{k=1}^{2j+1} \sum_{m=0}^k \frac{(-1)^m}{k} \frac{1}{2^{(k-m)} m! (2j-k+1)!} \left(\left(\frac{k-m}{2}\right)!\right)^{-2} \frac{(1+(-1)^{(k-m)})}{2} \left(\frac{1}{k} + \left(\psi_{\left(\left(\frac{k-m}{2}\right)+1\right)} - \psi_1\right)\right)$$

(3.9)

$$f_{10}(x) = \int_x^\infty \frac{1}{y} \int_0^\infty e^{-y\cosh(w)} \frac{1}{\cosh(w)} \, dw \, dy =$$

$$= -\pi \ln(2) - \frac{\pi}{2}(\gamma + \ln(x/2)) - (\gamma + \ln(x/2)) \sum_{j=0}^\infty \frac{1}{(2j+1)^2 \, 2^{2j} \, (j!)^2} x^{2j+1}$$

$$+ \sum_{j=0}^\infty \frac{1}{(2j+1)^3 \, 2^{2j-1} \, (j!)^2} x^{2j+1} + \sum_{j=1}^\infty \frac{1}{(2j+1)^2 \, 2^{2j} \, (j!)^2} \left(\psi_{(j+1)} - \psi_1\right) x^{2j+1}$$

(3.10)

$$f_{11}(x) = \int_x^\infty \frac{1}{y} \int_0^\infty e^{-y\cosh(w)} \ln\left(\frac{\cosh(w)}{\sinh(w)}\right) dw \, dy =$$

$$= -\frac{7}{8}\zeta_{(3)} - \frac{\pi^2}{8}(\gamma + \ln(x/2)) - \frac{\pi^2}{8} \sum_{j=1}^\infty \frac{1}{2^{2j+1} \, j \, (j!)^2} x^{2j} + \frac{\pi}{2} \sum_{j=0}^\infty \frac{1}{(2j+1)((2j+1)!!)^2} x^{2j+1}$$

$$- (\gamma + \ln(x/2)) \sum_{j=0}^\infty \frac{1}{(2j+2)} c_{1,j} \, x^{2j+2} + \sum_{j=0}^\infty \frac{1}{(2j+2)^2} c_{1,j} \, x^{2j+2} - \sum_{j=0}^\infty \frac{1}{(2j+2)} c_{2,j} \, x^{2j+2}$$

$$c_{1,j} = -\frac{1}{4} \sum_{m=0}^j \frac{(2j - 4m + 1)}{(2m+1)(j - m + 1) \, 2^{2m} \, (m!)^2 \, ((2j - 2m + 1)!!)^2}$$

$$c_{2,j} = \frac{1}{2} \sum_{m=0}^j \frac{1}{(2m+1) \, 2^{2m} \, (m!)^2 \, ((2j - 2m + 1)!!)^2} \left(\frac{1}{(2m+1)} + \frac{(2j - 4m + 1)}{2(j - m + 1)} \left(\psi_{(m+1)} - \psi_1\right)\right)$$

(3.11)

$$f_{12}(x) = \int_x^\infty \frac{1}{y} \int_0^\infty e^{-y\cosh(w)} \cosh(w) \frac{w}{\sinh(w)} \, dw \, dy =$$

$$= -\frac{\pi^2}{8}\ln(2) - \frac{25}{48}\zeta_{(3)} - \frac{1}{6}(\gamma + \ln(x/2))^3 - \frac{\pi^2}{6}(\gamma + \ln(x/2)) - \frac{1}{4}(\gamma + \ln(x/2))^2 \sum_{j=1}^\infty \frac{1}{j \, (2j)!} x^{2j}$$

$$- \frac{\pi^2}{12} \sum_{j=1}^\infty \frac{1}{j \, (2j)!} x^{2j} + \frac{\pi^2}{4} \sum_{j=0}^\infty \frac{1}{(2j+1)(2j+1)!} x^{2j+1}$$

$$- (\gamma + \ln(x/2)) \sum_{j=1}^\infty \left(\frac{(2j+1)}{2j} r_{1,j} - \frac{1}{(2j)^2 (2j+1)!}\right) x^{2j}$$

$$+ \sum_{j=1}^\infty \left(\frac{1}{(2j)^2} r_{1,j} + \frac{(2j+1)}{2j} r_{2,j} - \frac{1}{(2j)^3 (2j+1)!}\right) x^{2j}$$

$$r_{1,j} = \sum_{k=1}^{2j+1} \sum_{m=0}^k \frac{(-1)^m}{k} \frac{1}{2^{(k-m)} \, m! \, (2j - k + 1)!} \left(\left(\frac{k-m}{2}\right)!\right)^{-2} \frac{\left(1 + (-1)^{(k-m)}\right)}{2}$$

$$r_{2,j} = \sum_{k=1}^{2j+1} \sum_{m=0}^k \frac{(-1)^m}{k} \frac{1}{2^{(k-m)} \, m! \, (2j - k + 1)!} \left(\left(\frac{k-m}{2}\right)!\right)^{-2} \frac{\left(1 + (-1)^{(k-m)}\right)}{2} \left(\frac{1}{k} + \left(\psi_{\left(\left(\frac{k-m}{2}\right)+1\right)} - \psi_1\right)\right)$$

(3.12)

$$f13(x) = \int_x^\infty \frac{1}{y} \int_0^\infty e^{-y \cdot \cosh(w)} \, w \, \frac{\sinh(w)}{\cosh(w)} \, dw \, dy =$$

$$= -\frac{1}{12} \zeta_{(3)} - \frac{1}{6} (\gamma + \ln(x/2))^3 - \frac{\pi^2}{24} (\gamma + \ln(x/2)) - (\gamma + \ln(x/2)) \sum_{j=1}^\infty \frac{1}{2^{2j+2} j^2 (j!)^2} x^{2j}$$

$$+ \sum_{j=1}^\infty \frac{1}{2^{2j+2} j^3 (j!)^2} x^{2j} + \sum_{j=1}^\infty \frac{1}{2^{2j+2} j^2 (j!)^2} \left( \psi_{(j+1)} - \psi_1 \right) x^{2j}$$

(3.13)

$$f14(x) = \int_x^\infty \frac{1}{y} e^{-y} \, dy = -\gamma - \ln(x) - \sum_{j=1}^\infty \frac{1}{j! \, j} (-x)^j$$

(3.14)

$$f15(x) = e^{-x} = \sum_{j=0}^\infty \frac{1}{j!} (-x)^j$$

(3.15)

**Notes I:**

1) The functions $K_0(x)$ and $K_1(x)$ presented at the formulas for the functions f01(x), f02(x) and f03(x) represents respectivelly to the modified second-type Bessel functions of rank 0 and 1.

2) In the formulas above the symbol $\gamma$ represents the Euler-Mascheroni Gamma constant: $\gamma = 0.577215664901533...$

3) The symbol $\zeta(3)$ represents the Riemann Zeta function $\zeta(n)$ for n = 3: $\zeta(3) = 1.20205690315959...$

4) The terms $\left( \psi_{(j+1)} - \psi_1 \right)$ represents operations over the Euler Psi function with integer arguments:

$$\psi_1 = -\gamma \quad ; \quad \psi_{(j+1)} = -\gamma + \sum_{k=1}^j \frac{1}{k} \quad ; \quad \left( \psi_{(j+1)} - \psi_1 \right) = \sum_{k=1}^j \frac{1}{k}$$

**Notes II:**

In the resulting formula for $I_{KS}(x)$, in (3.3), we omitted the explicit mention to a 16th and to a 17th functions because the results for these two functions can be written in terms of the zeroth rank modified second-type Bessel function $K_0(x)$. The polynomial coeficients associated to both these functions, f16(x) and f17(x), were added to the polynomial coeficient that multiplies to f02(x). Nevertheless, it is worth to mention the interesting formulas for these two functions:

$$f16(x) = \int_0^\infty e^{-x \cdot \cosh(w)} \, w \, \sinh(w) \, dw = \frac{K_0(x)}{x}$$

(3.16)

$$f17(x) = \int_0^\infty e^{-x \cdot \cosh(w)} \ln(\sinh(w)) \, dw = -\frac{1}{2} (\gamma + \ln 2 + \ln x) K_0(x)$$

(3.17)

# 4. Limits and approximate formulas for the Källén-Sabry integral, $I_{KS}(x)$

## 4.1 Approximate formula valid for $x < 1$

In the literature [1, 15-20] it is possible to find several approximate formulas for the vacuum polarization potentials, some of them obtained by the simplification of their integral formulas and another obtained by means of curve fittings to the numeric values obtained by computer calculation.

In ref. [1], Blomqvist presented the analytic approximate formulas for the Uehling, Källén-Sabry and Wichmann-Kroll vacuum polarization potentials for small *r* values calculated up to the order $r^3 log(r)$. Nevertheless, due to the slow convergence of the power series expansions for these potential formulas they provide good approximations only for $r \ll 1$ and, in order to obtain a more precise approximation, it is necessary to add more terms with higher *r* powers.

The approximate formula for $I_{KS}(x)$ shown bellow was calculated using the exact power series obtained in section 3. The inclusion of terms with higher *r* powers allows this expansion to provide values with deviations smaller than 0.37% for $x < 1$, smaller than 0.19% for $x < 0.5$ and smaller than 0.0003% for $x < 0.1$

$$
\begin{aligned}
I_{KS}(x) \simeq & \left(-\zeta_{(3)} - \frac{65}{648} - \frac{\pi^2}{27}\right) - \frac{4}{9}(\gamma + \ln(x/2))^2 - \frac{13}{54}(\gamma + \ln(x/2)) \\
& + \left(\frac{13\pi^2}{18} + \frac{16}{9}\pi\ln(2) - \frac{383\pi}{135}\right)x + \frac{5}{12}x^2(\gamma + \ln(x/2)) - \frac{65x^2}{72} + \left(\frac{7\pi^2}{108} - \frac{10\pi}{81}\right)x^3 \\
& - \frac{5}{288}x^4(\gamma + \ln(x/2))^2 + \frac{4187}{86400}x^4(\gamma + \ln(x/2)) + \left(\frac{\zeta_{(3)}}{96} - \frac{33841}{207360} - \frac{5\pi^2}{3456}\right)x^4 \\
& + \left(\frac{11\pi}{3375} - \frac{\pi^2}{3600}\right)x^5 - \frac{7}{17280}x^6(\gamma + \ln(x/2))^2 + \frac{17833}{7776000}x^6(\gamma + \ln(x/2)) \\
& + \left(-\frac{1822711}{466560000} - \frac{7\pi^2}{207360}\right)x^6 + \left(\frac{34\pi}{2083725} + \frac{\pi^2}{127008}\right)x^7 - \frac{29}{4423680}x^8(\gamma + \ln(x/2))^2 \\
& + \frac{32429}{371589120}x^8(\gamma + \ln(x/2)) + \left(-\frac{13296077}{195084288000} - \frac{29\pi^2}{53084160}\right)x^8 \\
& + \left(\frac{23\pi}{241116750} + \frac{11\pi^2}{97977600}\right)x^9 - \frac{167}{2985984000}x^{10}(\gamma + \ln(x/2))^2 \\
& - \frac{43214671}{144850083840000}x^{10}(\gamma + \ln(x/2)) + \left(\frac{26826183589}{58403553804288000} - \frac{167\pi^2}{35831808000}\right)x^{10} \\
& + \left(\frac{58\pi}{124804708875} + \frac{17\pi^2}{18441561600}\right)x^{11} + \ldots
\end{aligned}
$$
(4.1.1)

## 4.2 Approximate formula for $x \geq 7$ and limit for $x \to +\infty$

It can be easily shown that the $I_{KS}(x)$ limit for $x \to +\infty$ is :

$$\lim_{x \to \infty} I_{KS}(x) = -\frac{\pi^2}{2}\frac{e^{-x}}{x}$$
(4.2.1)

The above formula is valid only for $x \gg 1$ and this limitation makes it unsuitable to provide usefull results for x values into the interval $7 > x > 0$ ($3.5\,\lambda_e > r > 0$) defined by Huang [16]. The reason for this is the slow convergence of the power series expansion for $I_{ks}(x)$ as a function of $1/x$ and $(1/x)^{1/2}$ for $x > 1$.

It is possible to introduce a small improvement into (4.2.1) by adding the first order term for the expansion of $I_{ks}(x)$ in terms of $1/x$ and $(1/x)^{1/2}$, thus obtaining :

$$I_{KS}(x) \simeq -\frac{\pi^2}{2}\frac{e^{-x}}{x} + \left(3\ln[2] - \frac{2}{9}\right)\sqrt{\frac{\pi}{2}}\frac{e^{-x}}{x^{3/2}}$$
(4.2.2)

Despite the formulas (4.2.1) and (4.2.2) being able to provide a good agreement with the order of magnitude for the exact $I_{KS}(x)$ values calculated using the formulas (3.0)-(3.15) into the interval $4 < x < 12$, the percentual difference between the values obtained using (4.2.2) and the exact values for $I_{KS}(x)$ represents 79% for x=4 and 51% for x=9.

### 4.3 Fast convergent formula for computational evaluation of $I_{KS}(x)$

By applying to eq. (2.2) the same change of variables ($t \rightarrow cosh(w)$) described in section 2 it is possible to change its formula in a way it becomes more suitable for the computational evaluation of $I_{KS}(x)$. A further calculation can turn the second integral contained in $I_{KS}(x)$ into a power series expansion, providing an even simpler and faster convergent single integral expression for the computational calculation of $I_{KS}(x)$.

$$
\begin{aligned}
I_{KS}(x) = \int_0^\infty e^{-x \cdot \cosh(w)} &\left\{ \left( \frac{13}{54} - \frac{19}{108} (\cosh(w))^{-2} + \frac{17}{108} (\cosh(w))^{-4} - \frac{2}{9} (\cosh(w))^{-6} \right) \right. \\
&+ \left( -\frac{44}{9} (\cosh(w))^{-1} + \frac{2}{3} (\cosh(w))^{-3} + \frac{5}{4} (\cosh(w))^{-5} + \frac{2}{9} (\cosh(w))^{-7} \right) w \sinh(w) \\
&+ 2 \ln(2) \left( 2 - (\cosh(w))^{-2} - (\cosh(w))^{-4} \right) + \frac{2}{3} \left( 2 - (\cosh(w))^{-2} - (\cosh(w))^{-4} \right) \ln\left( \frac{\cosh(w)}{\sinh(w)} \right) \\
&+ 2 \left( 2 - (\cosh(w))^{-2} - (\cosh(w))^{-4} \right) \ln(\sinh(w)) \\
&+ \frac{2}{3} \left( -4 (\cosh(w))^{-1} + (\cosh(w))^{-5} \right) \sinh(w) \left( -2 w \ln(1 - e^{-2w}) - w \ln(1 + e^{-2w}) + 2 \sum_{k=1}^\infty \frac{1}{k^2} e^{-2kw} + \sum_{k=1}^\infty \frac{(-1)^k}{k^2} e^{-2kw} \right) \right\} dw
\end{aligned}
$$

(4.3.1)

# 5. Table of values of the Källén-Sabry integral, $I_{KS}(x)$

| x | $I_{KS}(x)$ | x | $I_{KS}(x)$ | x | $I_{KS}(x)$ |
|---:|---:|---:|---:|---:|---:|
| 0,00001 | -58,9707080192 | 1,0 | -0,4249491222 | 4,8 | -0,0038677825 |
| 0,00005 | -43,8729457524 | 1,1 | -0,3652842419 | 4,9 | -0,0034541442 |
| 0,0001 | -38,0799719431 | 1,2 | -0,3153515711 | 5,0 | -0,0030854224 |
| 0,0002 | -32,7139632516 | 1,3 | -0,2732261771 | 5,1 | -0,0027566458 |
| 0,0003 | -29,7729402348 | 1,4 | -0,2374517236 | 5,2 | -0,0024634071 |
| 0,0004 | -27,7748156177 | 1,5 | -0,2069029108 | 5,3 | -0,0022017961 |
| 0,0005 | -26,2755641695 | 1,6 | -0,1806948257 | 5,4 | -0,0019683428 |
| 0,0006 | -25,0834040753 | 1,7 | -0,1581212485 | 5,5 | -0,0017599661 |
| 0,0007 | -24,0984671613 | 1,8 | -0,1386114848 | 5,6 | -0,0015739288 |
| 0,0008 | -23,2623213727 | 1,9 | -0,1216994181 | 5,7 | -0,0014077988 |
| 0,0009 | -22,5379198539 | 2,0 | -0,1070008481 | 5,8 | -0,0012594138 |
| 0,001 | -21,9003478588 | 2,1 | -0,0941965809 | 5,9 | -0,0011268505 |
| 0,002 | -17,9511651767 | 2,2 | -0,0830196043 | 6,0 | -0,0010083982 |
| 0,003 | -15,8381628701 | 2,3 | -0,0732452113 | 6,1 | -0,0009025342 |
| 0,004 | -14,4269928635 | 2,4 | -0,0646832992 | 6,2 | -0,0008079027 |
| 0,005 | -13,3826074846 | 2,5 | -0,0571722903 | 6,3 | -0,0007232967 |
| 0,006 | -12,5617663496 | 2,6 | -0,0505742849 | 6,4 | -0,0006476406 |
| 0,007 | -11,8904947097 | 2,7 | -0,0447711577 | 6,5 | -0,0005799761 |
| 0,008 | -11,3258165391 | 2,8 | -0,0396613898 | 6,6 | -0,0005194492 |
| 0,009 | -10,8406544093 | 2,9 | -0,0351574765 | 6,7 | -0,0004652983 |
| 0,01 | -10,4169006068 | 3,0 | -0,0311837916 | 6,8 | -0,0004168441 |
| 0,02 | -7,8683665824 | 3,1 | -0,0276748181 | 6,9 | -0,0003734807 |
| 0,03 | -6,5675863828 | 3,2 | -0,0245736721 | 7,0 | -0,0003346678 |
| 0,04 | -5,7279158049 | 3,3 | -0,0218308679 | 7,1 | -0,0002999229 |
| 0,05 | -5,1232597823 | 3,4 | -0,0194032768 | 7,2 | -0,0002688152 |
| 0,06 | -4,6588999352 | 3,5 | -0,0172532483 | 7,3 | -0,0002409605 |
| 0,07 | -4,2867389781 | 3,6 | -0,0153478639 | 7,4 | -0,0002160151 |
| 0,08 | -3,9792433408 | 3,7 | -0,0136583017 | 7,5 | -0,0001936725 |
| 0,09 | -3,7192932002 | 3,8 | -0,0121592939 | 7,6 | -0,0001736587 |
| 0,1 | -3,4955773951 | 3,9 | -0,0108286616 | 7,7 | -0,0001557287 |
| 0,2 | -2,2217521316 | 4,0 | -0,0096469159 | 7,8 | -0,0001396635 |
| 0,3 | -1,6249537724 | 4,1 | -0,0085969132 | 7,9 | -0,0001252674 |
| 0,4 | -1,2624544511 | 4,2 | -0,0076635589 | 8,0 | -0,0001123655 |
| 0,5 | -1,0145210112 | 4,3 | -0,0068335501 | 8,5 | -0,0000653073 |
| 0,6 | -0,8330321561 | 4,4 | -0,0060951529 | 9,0 | -0,0000376718 |
| 0,7 | -0,6942731072 | 4,5 | -0,0054380099 | | |
| 0,8 | -0,5849590901 | 4,6 | -0,0048529708 | | |
| 0,9 | -0,4969564898 | 4,7 | -0,0043319468 | | |

The above listed values have been calculated using a routine written in MS Visual Studio and performs the calculations in double precision (16 figures). For this reason the values of $I_{KS}(x)$ have been calculated just up to x = 9 and the number of figures for the results for higher values of x have been kept equal to 10.

## 6. Final remarks

Despite the inconvenience associated to its large sized expression, the result presented in section 3 provides an exact solution for the Källén-Sabry vacuum polarization potential. This result fullfils an existing gap given that analytic aproximate expressions for $V_{KS}(r)$ were known just for $r \ll 1$ and $r \to +\infty$ [1].

A relevant characteristic of the $f_j(x)$ functions, (3.1)-(3.17), is that one can't simply consider them as unrelated and independent parts of a solution for eq. (2.2). These functions share similar characteristics to the modified second-type Bessel functions and could, in a certain way, be interpreted as solutions for higher instances of complexity for systems described by the modified Bessel equation.

# Appendix A   An exact power series expansion formula for the Uehling vacuum polarization potential

The Uehling vacuum polarization potential [1] corresponds to the $\alpha Z\alpha$ order correction for the potential generated by a static point charge Z ($e = 1$). Its formula corresponds to:

$$V_{Uehling}(r) = -\frac{2}{3}\frac{\alpha Z\alpha}{\pi r} I_{Uehling}(2r) \tag{A.1}$$

where :

$$I_{Uehling}(x) = \int_1^\infty e^{-x.t}\left(t^{-2} + \frac{1}{2}t^{-4}\right)(t^2-1)^{1/2}\,dt = \tag{A.2.a}$$

$$= \int_0^\infty e^{-x.\cosh(w)}\left(1 - \frac{1}{2}(\cosh(w))^{-2} - \frac{1}{2}(\cosh(w))^{-4}\right)dw \tag{A.2.b}$$

In eq. (A.1), the value for the electron charge is set $e = 1$ and the value for the distance radius "$r$" from the point charge is given in units of reduced electron Compton wavelenght : $\lambdabar_e = $ 386.159 fm

By means of an algebric procedure and using the recurrence relations obeyed by the Bessel functions and by the Bickley-Naylor functions [8, 9] it is possible to rewritte (A.2.b) as:

$$I_{Uehling}(x) = \frac{1}{12}(12+x^2)K_0(x) - \frac{1}{12}x(10+x^2)K_1(x) + \frac{1}{12}x(9+x^2)\int_x^\infty K_0(y)\,dy \tag{A.3}$$

where :

$$K_0(x) = \int_0^\infty e^{-x.\cosh(w)}\,dw = -(\gamma+\ln(x/2))\sum_{j=0}^\infty \frac{1}{2^{2j}(j!)^2}x^{2j} + \sum_{j=1}^\infty \frac{1}{2^{2j}(j!)^2}\left(\psi_{(j+1)}-\psi_1\right)x^{2j} \tag{A.4}$$

$$K_1(x) = -\frac{d}{dx}K_0(x) = \int_0^\infty e^{-x.\cosh(w)}\cosh(w)\,dw =$$
$$= \frac{1}{x} + (\gamma+\ln(x/2))\sum_{j=0}^\infty \frac{1}{2^{2j+1}j!(j+1)!}x^{2j+1} - \sum_{j=1}^\infty \frac{1}{2^{2j+2}j!(j+1)!}\left(\psi_{(j+1)}-\psi_1\right)x^{2j+1} - \sum_{j=0}^\infty \frac{1}{2^{2j+2}j!(j+1)!}\left(\psi_{(j+2)}-\psi_1\right)x^{2j+1} \tag{A.5}$$

$$\int_x^\infty K_0(y)\,dy = \int_0^\infty e^{-x.\cosh(w)}\frac{1}{\cosh(w)}\,dw =$$
$$= \frac{\pi}{2} + (\gamma+\ln(x/2))\sum_{j=0}^\infty \frac{1}{(2j+1)\,2^{2j}(j!)^2}x^{2j+1} - \sum_{j=0}^\infty \frac{1}{(2j+1)^2\,2^{2j}(j!)^2}x^{2j+1} - \sum_{j=1}^\infty \frac{1}{(2j+1)\,2^{2j}(j!)^2}\left(\psi_{(j+1)}-\psi_1\right)x^{2j+1} \tag{A.6}$$

**Notes:**

1) The functions $K_0(x)$ and $K_1(x)$ corresponds respectively to the modified second-type Bessel functions of rank 0 and 1.

2) In the formulas above the symbol $\gamma$ represents the Euler-Mascheroni Gamma constant: $\gamma = $ 0.577215664901533...

3) The terms $\left(\psi_{(j+1)}-\psi_1\right)$ represents operations over the Euler Psi function with integer arguments:

$$\psi_1 = -\gamma \quad ; \quad \psi_{(j+1)} = -\gamma + \sum_{k=1}^j \frac{1}{k} \quad ; \quad \left(\psi_{(j+1)}-\psi_1\right) = \sum_{k=1}^j \frac{1}{k}$$

The aproximate formula for the Uehling integral for small x values can be written as:

$$
\begin{aligned}
I_{\text{Uehling}}(x) \simeq & -\frac{5}{6} - (\gamma + \ln(x/2)) + \frac{3\pi x}{8} - \frac{3x^2}{8} + \frac{\pi x^3}{24} - \frac{7x^4}{192} + \frac{1}{64} x^4 (\gamma + \ln(x/2)) \\
& - \frac{127 x^6}{345\,600} + \frac{1}{5760} x^6 (\gamma + \ln(x/2)) - \frac{949 x^8}{240\,844\,800} + \frac{1}{573\,440} x^8 (\gamma + \ln(x/2)) \\
& - \frac{6079 x^{10}}{195\,084\,288\,000} + \frac{1}{77\,414\,400} x^{10} (\gamma + \ln(x/2)) - \frac{5053 x^{12}}{27\,748\,152\,115\,200} \\
& + \frac{1}{14\,014\,218\,240} x^{12} (\gamma + \ln(x/2)) + .\ ...
\end{aligned}
\qquad (A.7)
$$